\documentclass{article}
\usepackage{graphicx}
\title{Control software analysis, Part I\\
Open-loop properties}
\author{Eric Feron~\thanks{Dutton/Ducoffe Professor of Aerospace Software Engineering,
Georgia Institute of Technology. {\tt feron@gatech.edu}} \and Fernando Alegre~\thanks{College of Computing, Georgia Institute of Technology. {\tt fernando@gatech.edu}}}
\def\dsp{\displaystyle}
\def\reals{{\bf R}}
\begin{document}
\bibliographystyle{alpha}
\maketitle

{\bf Abstract:} As the digital world enters further into everyday life, questions are raised about the increasing challenges brought by the interaction of real-time software with physical devices. Many accidents and incidents encountered in areas as diverse as medical systems, transportation systems or weapon systems are ultimately attributed to ``software failures''. Since real-time software that interacts with physical systems might as well be called control software, the long litany of accidents due to real-time software failures might be taken as an equally long list of opportunities for control systems engineering. In this paper, we are interested only in run-time errors in those pieces of software that are a direct implementation of control system specifications: For well-defined and well-understood control architectures such as those present in standard textbooks on digital control systems, the current state of theoretical computer science is well-equipped enough to address and analyze control algorithms. It appears that a central element to these analyses is Lyapunov stability theory, which translate into invariant theory in computer implementations. 

\section{Introduction}
\subsection{Preliminaries: An example system}
\begin{figure}[htbp]  
   \hspace{0mm}
   \begin{center}
    \includegraphics[width=10cm]{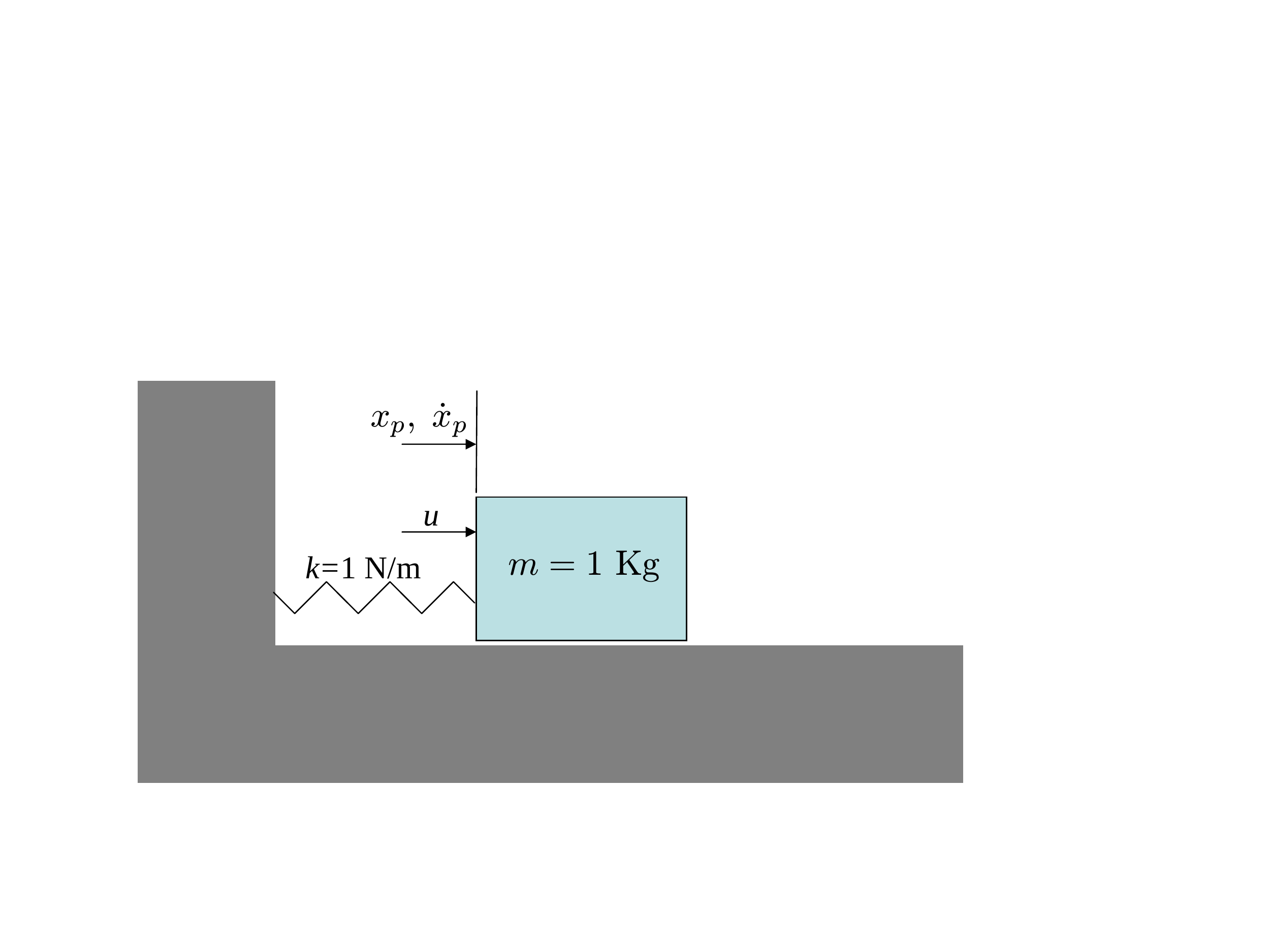}
    \end{center}
    \caption{Controlled physical system}
    \label{figure1}
\end{figure}

\begin{figure}[htbp]  
   \hspace{0mm}
   \begin{center}$\mbox{}$\vspace{-7cm}
    \includegraphics[width=14cm]{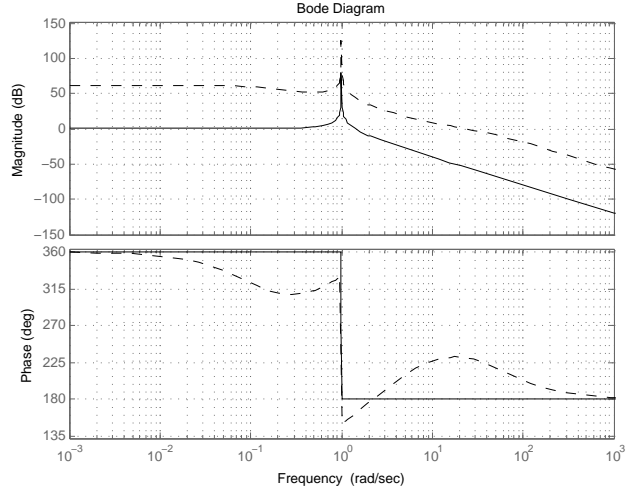}\vspace{-5cm}
    \end{center}
    \caption{Top: Bode plot. Continuous: loop gain of plant. Dashed: loop gain of plant + compensator.}
    \label{bode-step}
\end{figure}

\begin{figure}[htbp]  
   \hspace{0mm}
   \begin{center}
    \includegraphics[width=12cm]{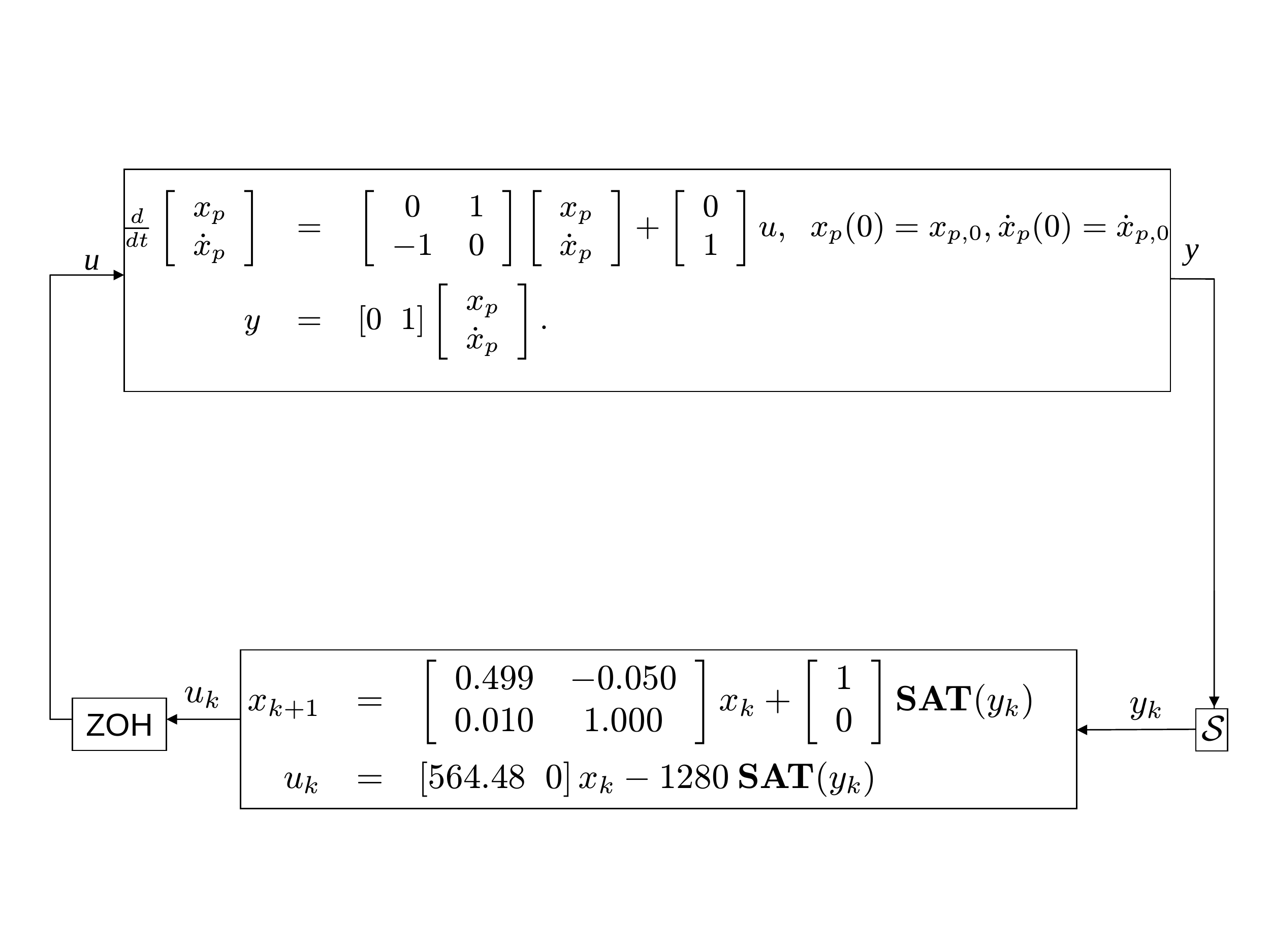}
    \end{center}
    \caption{Closed-loop system}
    \label{Figure2}
\end{figure}

We consider a standard control system, in this case a one-mass, one spring system. 
such as that shown in Fig.~\ref{figure1}.
The system is modeled via the following differential equation
\begin{equation}
\begin{array}{rcl}
\dsp\frac{d}{dt}\left[\begin{array}{c} x_p \\ \dot{x}_p \end{array}\right] &=& 
\left[\begin{array}{cc} 0 & 1 \\-1 & 0\end{array}\right] \left[\begin{array}{c}x_p \\ \dot{x}_p\end{array} \right]+ \left[\begin{array}{c}0 \\ 1 \end{array}\right]u,\;\; x(0) = x_0, \dot{x}_p(0) = \dot{x}_{p,0} \\ [10pt]
y & = & \left[0 \;\; 1\right]\left[\begin{array}{c}x_p \\ \dot{x}_p\end{array} \right].
\end{array}
\label{system}
\end{equation}
In these equations, $x_p$ and $\dot{x}_p$ are measured in meters and meters per second, respectively, while $u$ is expressed in Newtons. We assume that the output $y$ is available for feedback purposes. The chosen controller is of the lead-lag type, 
\begin{equation}
\begin{array}{l}
\tilde{y}(t) = {\bf SAT} (y(t)),\;\\
u(s) = \displaystyle -128\frac{s+1}{s+0.1}\frac{s/5+1}{s/50+1}\tilde{y}(s),
\end{array}
\label{compensator}
\end{equation}
which guarantees good closed-loop system bandwidth (in excess of 20 rad/sec) and good steady-state disturbance rejection properties; following common practices, the input to the controller $y$ is first passed through a saturation function to avoid variable overflow in the controller. 
The essential properties of the closed-loop system may be easily extracted from the Bode plot of the open-loop transfer function of the plant mounted in series with the controller, shown in Fig.~\ref{bode-step}. 
For example, the phase margin can be easily seen to exceed 50 degrees. This controller is then implemented on a computer by considering a sample-and-hold architecture running at 100 Hz, by implementing the state-space model
\begin{equation}
\begin{array}{rcl}
x_{k+1}&=&\left[\begin{array}{cc} 0.499 & -0.050 \\ 0.010 & 1.000\end{array} \right]x_{k} 
+\left[\begin{array}{c} 1 \\ 0 \end{array} \right] {\bf SAT}(y_k) \\
& = & Ax_{k} + B{\bf SAT}(y_k)\\[10pt]
u_{k} &=& \left[564.48 \;\; 0\right]x_{k} - 1280 \:{\bf SAT}(y_k)\\
&=&Cx_{k}+D{\bf SAT}(y_k)
\end{array}
\label{disc-controller}
\end{equation}
which is no more than a first-order Euler discretization of the following state-space realization of the controller~(\ref{compensator})
\[
\begin{array}{rcl}
\dsp \frac{d}{dt}x& = &\left[\begin{array}{cc}-50.1 & -5.0 \\1.0 & 0.0 \end{array}\right]x+ \left[\begin{array}{c}100\\0\end{array} \right]{\bf SAT}(y)\\[10pt]
u &=& \left[564.48 \;\; 0 \right] x - 1280 \:{\bf SAT}(y).
\end{array}
\]
In the expression for the compensator, $y_k$ is $y(0.01k)$, the output of the system sampled at time $0.01k$. Likewise, $u_k$ is the compensator output fed to the plant during the time interval $\left[ 0.01k \;\; 0.01(k+1)\right)$, resulting in the closed-loop description of the system as shown in Fig.~\ref{Figure2}. 
This figure includes the sampling process ${\cal S}$ and the zero-order hold ZOH.
An implementation of the compensator may then be realized by means of the simplified, but functional C code shown in Fig.~\ref{code}. 
\begin{figure}
\begin{center}
\begin{verbatim}
1: int main(int argc, char *argv[])
2: {
3:   double *x[2], *xb[2], y, u;
4:   x[0] = 0;
5:   x[1] = 0;
6:   while(1){
7:     fscanf(stdin,"%f",&y);
8:     if (y >1){
9:       y=1
10:	   }
11:	   if (y<-1){
12:      y=-1
13:	   }
14:    u = 564.48*x[0]-1280*y;
15:    fprintf(stdout,"%f\n",u);
16:    xb[0]=x[0];
17:    xb[1]=x[1];
18:    x[0]:= 0.4990*xb[0]-0.0500*xb[1]+y;
19:    x[1]:= 0.01*xb[0]+xb[1];
20:  }
21:}
\end{verbatim}
\end{center}
\caption{C implementation of controller}
\label{code}
\end{figure}
The final step involves the compilation of the C code to executable code, and the eventual operation of the system in closed-loop. 

\subsection{Performance and stability questions}
Considering the entire system, made of the physical plant and its controller, we are led to asking whether ``the system works''. In this paper, we are interested in asking this question about the software implementation of the control 
law shown in Fig.~\ref{code} and the corresponding executable binary code. 

Many control engineers and researchers may wonder whether this question makes any sense at all: After all, control system textbooks suggest that the task of verifying whether the closed-loop system ``works'' amounts to establishing, in some way or some other, that the closed-loop ``system'' that consists of the system~(\ref{system}) combined with either the continuous-time controller~(\ref{compensator}) or the discrete-time controller~(\ref{disc-controller}) is stable, which requires no more than computing the roots of a low-order polynomial. Further, control systems manufacturers may require the controller itself to be open-loop stable. Other questions about system performance, including disturbance rejection properties and robustness against modeling inaccuracies, may also be easily answered by using algebraic or graphical methods.

Note, however, that the process of generating C code from the continuous or discrete control laws is often done automatically, by means of commercial autocoders whose implementation details are not fully known. Likewise, proprietary or public-domain compilers are then used to produce binary executables from the C code. While autocoders and compilers can be assumed to operate in good faith, they cannot just be trusted blindly. In addition, the codes generated out of the specifications often contain detailed computations whose details are not known at the level of control law specifications. Thus the produced C code and the binary executable must be examined in detail for adequate performance by following a process that is {\em independent} from the original code generation process. The amount of work involved in such an endeavour may appear formidable. However, the alternative which is to certify the correctness 
of autocoders and compilers is even more daunting~\cite{Rin:99,RV2004ArkoudasAndRinard,rinard99credible}.

The rest of this paper is devoted to bringing forward the elements that make the static analysis of code that implements control system specification not only possible, but also necessary given how easy it is to perform.

\section{Code Analysis} 
\subsection{Performance analysis and Hoare logic}
Analyzing the software that implements a control system consists of answering several questions: 
\begin{itemize}
\item Does the control function keep executing properly and within acceptable execution time limits?
\item Is the implemented controller stable and what is the range of values explored by each variable?
\item Is the closed-loop system stable?
\item Does the software implement the specifications?
\end{itemize}
These questions may be extended to include further concerns about system performance. 
However, the problem remains fundamentally the same: That is, to 
reach the same level of confidence about system 
stability and performance when considering the controller code in Fig.~\ref{code}
together with the plant~(\ref{system})
as that reached when analyzing the controller 
specification~(\ref{disc-controller}), together with the 
plant~(\ref{system}). To reach that goal, two options exist: 

The first option is to follow a ``macroscopic pattern matching'' approach, whereby 
the controller specification~(\ref{disc-controller}) is reconstructed from the code in Fig.~\ref{code}. 
This solution, adopted in~\cite{CousotEtAl-ESOP05,Fer:04}, is particularly advantageous when the 
code structure is relatively repeatable from one instance to the other. However, pattern matching can become more challenging if codes that implement controllers do not have well-identified and rigid structures. 

The second option, which we call code-level analysis, is to develop proofs that directly identify the code semantics.
Code-level analysis is, in principle, supported by an elegant logical environment that closely matches available control oriented methodologies. This environment, developed by Floyd and Hoare in the late 1960s~\cite{Flo:67,Hoa:69}, 
intends to describe the code semantics by concatenating 
elementary knowledge about each line of code~\cite[Ch.7]{Pel:01}: Each line of code is annotated with one precondition {\em pre} satisfied by the program state prior to the line execution, and one postcondition {\em post} satisfied by the program state after the execution of the line. Thus each line of code {\tt loc} is transformed into a {\em Hoare triple} of the kind {\em \{pre\}}{\tt loc}{\em \{post \}}, where the {\em pre} and {\em post} are compatible with {\tt loc}. Global program properties can then be inferred by concatenating Hoare triples, as follows: Considering two consecutive Hoare triples {\em \{pre\_1\}}{\tt loc\_1}{\em \{post\_1 \}} and {\em \{pre\_2\}}{\tt loc\_2}{\em \{post\_2 \}} and the property {\em \{post\_1\} $\rightarrow$ \{pre\_2\}}, whe have 
{\em \{pre\_1\}}{\tt loc\_1}, {\tt loc\_2}{\em \{post\_2 \}}. Further developments of Hoare logic will be used thereafter.

\subsection{Hoare logic and Lyapunov functions}

We now examine how Hoare logic can be used to formalize the analysis of control system stability requirements for the control system specifications and their implementation.  Informally speaking, we know that the system that implements the controller is stable and all states are bounded because the matrix 
\[
A=\left[\begin{array}{cc} 0.499 & -0.050 \\ 0.010 & 1.000\end{array} \right]
\]
is stable, that is, its two eigenvalues have modulus less than one, and the input is effectively bounded in absolute value by one. While exact, this proof 
does not carry over easily to the software implementation. 
Moreover, it is somewhat incomplete since it does not 
tell anything related to the range covered by the variables involved. These issues may 
be addressed by means of Lyapunov stability theory: Consider the Lyapunov function $V(x)=x^TPx$, where 
\[
{P} = \left[\begin{array}{cc} 0.0300  &  0.2000\\
0.2000 &  10.0000 \end{array}\right]
\]
and $P$ is a positive definite matrix.
Then it is easy to show that the value of $x^TPx$ decays along the trajectories of 
\[
x_{k+1} = A x_k, 
\]
that is $V(x_k+1)-V(x_k)<0$ if $y_k = 0$ and $x_k \neq 0$. That alone indicates stability of the controller, yet it does not bring any indication about the size of the states of the system. For that purpose, and following well-known practices~\cite{BEFB:94}, it is possible to show that $P$
defines not only a valid Lyapunov function, but also an invariant ellipsoid 
\[
{\cal E}_{P} = \left\{ x \in \reals^2 \; |\; x^TPx \leq 1 \right\}.
\]
for any input $y$, and this ellipsoid defines a bounded region for $x_k$. In particular, if $x_0$ is included in ${\cal E}_{{P}}$ (and it does since $x_0=0$), then the state will stay in the ellipsoid forever. A simple eigenvalue computation on $P$ shows that each entry of the state $x$ is bounded in absolute value by 7.

Let us now examine how this knowledge can be included in the controller's executable code so as to establish the code's correctness. We may begin our effort with analyzing a Matlab script that implements the controller, given in Fig.~\ref{matlab-code}. The reason for doing so is that a Matlab code structure is sufficiently close to typical control system specifications to make the correspondence between traditional proofs and code proof transparent, while remaining executable by Matlab's interpreter.
\begin{figure}
\begin{verbatim}
1:  A = [0.4990, -0.0500; 0.0100, 1.0000];
2:  C = [564.48, 0];
3:  B = [1;0];D = -1280;
4:  x = zeros(2,1);
5:  while 1
6:   y = fscanf(stdin,"%f");
7:   y = max(min(y,1),-1);
8:   u = C*x + D*y;
9:   fprintf(stdout,"%f\n",u);
10:   x = A*x + B*y;
11: end
\end{verbatim}
\caption{MATLAB implementation of controller}
\label{matlab-code}
\end{figure}
Identifying $x$ as the controller's state, we now proceed with establishing the stability of 
the controller's Matlab implementation by using Hoare logic. For that, we may begin the 
program annotation at any line of choice, for example at line {\tt 8: u = C*x + Dy}. This assignment does not influence the state $x$ but determines the value of $u$. It also introduces the controller input $y$. Thus we propose the following triple:

\begin{equation}
\begin{array}{l}
\left\{x \in {\cal E}_{{P}} ,\; y^2 \leq 1 \right\} \\
\mbox{\tt 8: u = C*x+D*y;}\\
\left\{x \in {\cal E}_{{P}},\;y^2\leq 1, \; u^2 \leq 2(C{P}^{-1}C^T+ D^2) \right\}
\end{array}
\label{triple}
\end{equation}

Much about this statement may appear to be rather arbitrary at first sight, and indeed requires 
global knowledge of the program (or its specifications, described earlier) to be established. However, the statement $\left\{u^2 \leq 2(C{P}^{-1}C^T+ D^2)\right\}$ can be easily figured out from the value of $u$ and the statements $\left\{x \in {\cal E}_{{P}} ,\; y^2 \leq 1 \right\}$. From there on, it is possible to obtain statements about the evolution of the state-space by means of backwards propagation~\cite{Pel:01}. To begin with, line {\tt 7: y = max(min(y,1),-1)} implicitly indicates that $y^2 \leq 1$ after this assignment, and can be commented via the triple

\vspace{3mm}
\begin{tabular}{l}
\{$x \in {\cal E}_{{P}}$\}\\
{\tt 7: y = max(min(y,1),-1);}\\
$\left\{x \in {\cal E}_{{P}},\; y^2 \leq 1  \right\}$
\end{tabular}
\vspace{3mm}

At this point,  the assignment in line {\tt 6: y = fscanf(stdin,"\%f")} can simply 
be surrounded by the same statement $\left\{x \in {\cal E}_{{P}}\right\}$ to yield
the triple

\vspace{3mm}
\begin{tabular}{l}
$\{x \in {\cal E}_{{P}}\}$\\
{\tt 6: y = fscanf(stdin,"\%f")}\\
$\{x \in {\cal E}_{{P}}\}$
\end{tabular}
\vspace{3mm}

Line {\tt 5: while 1} may not be separated from line {\tt 11: end}. Following standard Hoare logic~\cite{Pel:01,Mon:03}, these two lines may be documented as follows:

\vspace{3mm}
\begin{tabular}{l}
$\{x \in {\cal E}_P\}$\\
{\tt 5:  while 1}\\
$\{x \in {\cal E}_P\}$\\
\vdots \\
$\{x \in {\cal E}_P\}$\\
{\tt 11: end}\\
$\{ false \}$
\end{tabular}
\vspace{3mm}

Moving backwards, we see that the two lines that must be worked on are lines 4 and 10. 
Considering first the assignment {\tt 4: x = zeros(2,1)}, we are led to the triple

\vspace{3mm}
\begin{tabular}{l}
$\{ true \}$\\ 
{\tt 4: x = zeros(2,1)}\\
$\{x \in {\cal E}_P\}$.
\end{tabular}
\vspace{3mm}

Moving further backwards 
yields no changes concerning the state variable $x$, and triples may be constructed for lines 1 through 3 
according to the template

\vspace{3mm}
\begin{tabular}{l}
$\{ true \}$\\ 
{\tt loc}\\
$\{true \}$.
\end{tabular}
\vspace{3mm}


Line {\tt 10:   x = A*x + B*y} introduces the controller input $y$ and performs an important transformation on the state $x$. Using insight from line {\tt 7:   y = max(min(y,1),-1)} we introduce the condition $y^2\leq 1$. 
We also propagate the statement $\left\{x \in {\cal E}_{{P}}  \right\}$
backwards: Replacing $x$ by its value $Ax+B y$ yields 
the statement $\left\{Ax + B y \in {\cal E}_{P} \right\}$. Thus the triple 

\vspace{3mm}
\begin{tabular}{l}
$\left\{Ax+B y \in {\cal E}_{{P}}, \; y^2 \leq 1\right\}$\\
{\tt 10: x = A*x + B*y;} \\
$\left\{x \in {\cal E}_{{P}}  \right\}$
\end{tabular}
\vspace{3mm}

Moving to line {\tt 9: fprintf(stdout,"\%f$\backslash$n",u)}, we see that it really does not act in any way on the state of the program. However, it introduces the controller output value $u$, whose numerical range must be known. At this point, we {postulate} the property $u^2 \leq 2(CP^{-1}C^T + D^2)$ and we are led to the triple

\vspace{3mm}
\begin{tabular}{l}
$\{Ax+By \in {\cal E}_P, \; u^2 \leq 2(CP^{-1}C^T + D^2),\;y^2 \leq 1 \}$\\
{\tt 9:  fprintf(stdout,"\%f$\backslash$n",u)}\\
$\{Ax+By \in {\cal E}_P, \; y^2 \leq 1\}$.
\end{tabular}
\vspace{3mm}

For this triple to be ``compatible'' with the triple~(\ref{triple}) associated with line 8, we must prove 
the statement 
\begin{equation}
\{x \in {\cal E}_P, \; y^2 \leq 1\} \Rightarrow \left\{Ax+B y \in {\cal E}_{{P}}, \; y^2 \leq 1\right\}.
\label{quad_const}
\end{equation}
For the specific instances of $A$, $B$ and $P$ under consideration, this statement is indeed true, yet not 
trivial. A simple proof relies on the ${\cal S}$-procedure~\cite{AiG:64}, which amounts to introducing the auxiliary (and true) statement
\begin{equation}
\forall (x,y)\; (Ax+By)^TP(Ax+By)- 0.01 x^TPx - 0.99 y^2 \leq 0. 
\label{eqn2}
\end{equation}
leading to 
\[
\{x \in {\cal E}_P, \; y^2 \leq 1, \;(Ax+By)^TP(Ax+By)- 0.01 x^TPx - 0.99 y^2 \leq 0\} \Rightarrow \left\{Ax+B y \in {\cal E}_{{P}}, \; y^2 \leq 1\right\},
\]
which may now be checked much more easily than~(\ref{quad_const}) alone. Thus a preferrable triple for line 8 is 

\vspace{3mm}
\begin{tabular}{l}
$\left\{x \in {\cal E}_{{P}} ,\; y^2 \leq 1 \right\}$ \\
{\tt 8: u = C*x+D*y;}\\
$\left\{x \in {\cal E}_{{P}},\;y^2\leq 1, \right. $\\
$\left. (Ax+By)^TP(Ax+By)- 0.01 x^TPx - 0.99 y^2 \leq 0,\; u^2 \leq 2(C{P}^{-1}C^T+ D^2) \right\}$
\end{tabular}
\vspace{3mm}

Concatenating statements about all individual instructions together leads us to the following 
annotated Matlab code:

\vspace{3mm}
\begin{tabular}{l}
$\{true\}$\\
{\tt 1:  A = [0.4990, -0.0500; 0.0100, 1.0000];}\\
$\{true\}$\\
{\tt 2:  C = [-564.48, 0];}\\
$\{true\}$\\
{\tt 3: B = [1;0];D=1280}\\
$\{true\}$\\
{\tt 4: x = zeros(2,1);}\\
$\left\{x \in {\cal E}_{{P}}\right\}$\\
{\tt 5: while 1}\\
$\{x \in {\cal E}_{{P}}\}$\\
{\tt 6: y = fscanf(stdin,"\%f")}\\
$\{x \in {\cal E}_{{P}}\}$\\
{\tt 7: y = max(min(y,1),-1);}\\
$\left\{x \in {\cal E}_{{P}} ,\; y^2 \leq 1 \right\}$\\
{\tt 8: u = C*x+D*y;}\\
$\left\{x \in {\cal E}_{{P}},\;y^2\leq 1, \right. $\\
$\left. (Ax+By)^TP(Ax+By)- 0.01 x^TPx - 0.99 y^2 \leq 0,\; u^2 \leq 2(C{P}^{-1}C^T+ D^2) \right\}$\\
{\tt skip}\\
$\{Ax+By \in {\cal E}_P, \; u^2 \leq 2(CP^{-1}C^T + D^2),\;y^2 \leq 1 \}$\\
{\tt 9:  fprintf(stdout,"\%f$\backslash$n",u)}\\
$\left\{Ax+B y \in {\cal E}_{{P}}, \; y^2 \leq 1\right\}$\\
{\tt 10: x = A*x + B*y;} \\
$\left\{x \in {\cal E}_{{P}}  \right\}$\\
{\tt 11: end}\\
$\{false\}$
\end{tabular}
\vspace{3mm}

This annotated program contains the instruction $skip$ whose purpose is only to separate
two consecutive logical statements. 

\section{Forward vs. backwards propagation of constraints}
Although the backwards constraint propagation described in the previous section is easy to do and best matches the well-known procedures used to establish stability properties of dynamical systems, forward constraint propagation provides key insights about how these constraints may otherwise be formulated.

We now reconsider the Matlab code shown in Fig.~\ref{matlab-code}, with a forward exploration perspective. Once again, ellipsoids are used to bound all state variables and inputs involved in this program. Forward propagation 
allows us to begin at line {\tt 1: A = [0.4990, -0.0500; 0.0100, 1.0000]}, which can simply be surrounded by $\{true\}$ statements to produce the Hoare triple

\vspace{3mm}
\begin{tabular}{l}
$\{true\}$\\
{\tt 1:  A = [0.4990, -0.0500; 0.0100, 1.0000];}\\
$\{true\}$
\end{tabular}
\vspace{3mm}

Lines 2 and 3 can be treated similarly. Line {\tt 4: x = zeros(2,1)} introduces the variable $x$, and since $x = 0$, 
line 4 can be followed by the post-condition $\{x \in {\cal E}_P\}$ to yield the triple

\vspace{3mm}
\begin{tabular}{l}
$\{ true\}$\\
{\tt 4:  x = zeros(2,1);}\\
$\{ x \in {\cal E}_P\}$
\end{tabular}
\vspace{3mm}

\noindent which is the same as that given in the backwards analysis. Lines 5 and 11 introduce the controller loop,
and we {postulate} the pre-condition $\{ x \in {\cal E}_P\}$, which then leads to commented set of commands:

\vspace{3mm}
\begin{tabular}{l}
$\{x \in {\cal E}_P\}$\\
{\tt 5: while 1}\\
$\{ x \in {\cal E}_P\}$\\
$\vdots$ \\
$\{x \in {\cal E}_P\}$\\
{\tt 11: end}\\
$\{false\}$
\end{tabular}
\vspace{3mm}

Line {\tt 6: y = fscanf(stdin,"\%f");} merely introduces the variable $y$ with no particular property associated with it. Thus the triple

\vspace{3mm}
\begin{tabular}{l}
$\{x \in {\cal E}_P\}$\\
{\tt 6: y = fscanf(stdin,"\%f");}\\
$\{ x \in {\cal E}_P\}$
\end{tabular}
\vspace{3mm}

The next line {\tt 7:   y = max(min(y,1),-1);} is much more informative since it forces $y$ to be bounded by one in absolute value. This information may be encoded in various ways. One is to incorporate this information {\em verbatim}, therefore leading to the triple

\vspace{3mm}
\begin{tabular}{l}
$\{ x \in {\cal E}_P\}$\\
{\tt 7:   y = max(min(y,1),-1);}\\
$\{ x \in {\cal E}_P, \; y^2 \leq 1\}$
\end{tabular}
\vspace{3mm}

However, the postcondition $\{ x \in {\cal E}_P, \; y^2 \leq 1\}$ will not be used as the preconditon for line 
{\tt 8:   u = C*x + D*y;}. Instead, we introduce the weaker pre-condition 
\[
\left\{\left[\begin{array}{l}x\\y\end{array}\right] \in {\cal E}_Q \right\}, \;\; \mbox{ with } 
Q = \left[\begin{array}{cc} 0.01 P & 0 \\ 0 & 0.99 \end{array} \right].
\]
Asking what the precondition becomes after passing through line 8
is equivalent to asking what the image of an ellipsoid through a linear mapping is. 
Since it is also an ellipsoid (in this case 1-dimensional), 
we conclude that a valid clause in the postcondition of line 8 is
$u^2 \leq \left[C \; D \right]Q^{-1}\left[C\; D\right]^T$, while keeping the other clausal elements 
identical. The resulting triple is therefore

\vspace{3mm}
\begin{tabular}{l}
$\left\{\left[\begin{array}{l}x\\y\end{array}\right] \in {\cal E}_Q\right\}$\\
{\tt 8:   u = C*x + D*y;}\\
$\left\{\left[\begin{array}{l}x\\y\end{array}\right] \in {\cal E}_Q, \; 
u^2 \leq \left[C \; D \right]Q^{-1}\left[C\; D\right]^T\right\}$
\end{tabular}
\vspace{3mm}

\noindent which propagates the set to which $x$ and $y$ belong, while providing a bound on the output variable $u$.
Line {\tt 9:   fprintf(stdout,"\%f$\backslash$n",u);} merely writes the output of the controller to the appropriate memory location for further processing. Since $u$ is not used afterwards, its range may be dropped from the postcondition, 
thus leading to the triple

\vspace{3mm}
\begin{tabular}{l}
$\left\{\left[\begin{array}{l}x\\y\end{array}\right] \in {\cal E}_Q, \; u^2 \leq \left[C \; D \right]Q^{-1}\left[C\; D\right]^T \right\}$\\
{\tt 9:   fprintf(stdout,"\%f$\backslash$n",u);}\\
$\left\{\left[\begin{array}{l}x\\y\end{array}\right] \in {\cal E}_Q\right\}$
\end{tabular}
\vspace{3mm}

We finally consider line {\tt 10: x = A*x + B*y}. Again, this line transforms the ellipsoid ${\cal E}_Q$ into the 
ellipsoid ${\cal E}_{\tilde{P}}$, where
\begin{equation}
\tilde{P} = \left(\frac{1}{0.01} AP^{-1}A^T+\frac{1}{0.99}BB^T\right)^{-1}.
\label{ptilde}
\end{equation}
To conclude, thus line 10 leads to the following triple

\vspace{3mm}
\begin{tabular}{l}
$\left\{\left[\begin{array}{l}x\\y\end{array}\right] \in {\cal E}_Q\right\}$\\
{\tt 10:   x = A*x + B*y;}\\
$\{ x \in {\cal E}_{\tilde{P}}\}$
\end{tabular}
\vspace{3mm}

The documentation of the program by means of Hoare triple is now complete. It is necessary, however, to check that 
the post condition of line 10 implies the precondition of line 11, that is:
\[
\{ x \in {\cal E}_{\tilde{P}}\} \Rightarrow \{x \in {\cal E}_{P}\}.
\]
This is equivalent to showing that 
\begin{equation}
\left(\frac{1}{0.01} AP^{-1}A^T+\frac{1}{0.99}BB^T\right)^{-1} - P \leq 0,
\label{eqn1}
\end{equation}

\noindent which is indeed true for the given values of $A$, $P$ and $B$. Moreover, it can be shown that the clause~(\ref{eqn1}) is equivalent to~(\ref{eqn2}) for generic values of $A$, $B$, and $P$, using standard matrix manipulations~\cite{BEFB:94}, thus demonstrating the ``equivalence'' of the forward and backwards approaches to analyze this program. The forward analysis eventually yields the annotated program

\vspace{3mm}
\begin{tabular}{l}
$\{true\}$\\
{\tt 1:  A = [0.4990, -0.0500; 0.0100, 1.0000];}\\
$\{true\}$\\
{\tt 2:  C = [-564.48, 0];}\\
$\{true\}$\\
{\tt 3: B = [1;0];D=1280}\\
$\{true\}$\\
{\tt 4: x = zeros(2,1);}\\
$\left\{x \in {\cal E}_{{P}}\right\}$\\
{\tt 5: while 1}\\
$\{x \in {\cal E}_{{P}}\}$\\
{\tt 6: y = fscanf(stdin,"\%f")}\\
$\{x \in {\cal E}_{{P}}\}$\\
{\tt 7: y = max(min(y,1),-1);}\\
$\{ x \in {\cal E}_P, \; y^2 \leq 1\}$\\
skip\\
$\left\{\left[\begin{array}{l}x\\y\end{array}\right] \in {\cal E}_Q\right\}$\\
{\tt 8: u = C*x+D*y;}\\
$\left\{\left[\begin{array}{l}x\\y\end{array}\right] \in {\cal E}_Q, \; u^2 \leq \left[C \; D \right]Q^{-1}\left[C\; D\right]^T \right\}$\\
{\tt 9:  fprintf(stdout,"\%f$\backslash$n",u)}\\
$\left\{\left[\begin{array}{l}x\\y\end{array}\right] \in {\cal E}_Q\right\}$\\
{\tt 10: x = A*x + B*y;} \\
$\{ x \in {\cal E}_{\tilde{P}}\}$\\
{skip}\\
$\{x\in {\cal E}_P\}$\\
{\tt 11: end}
\end{tabular}
\vspace{3mm}

\section{{C} code analysis}
The latter analysis may now be extended to the original C code given in 
Fig.~\ref{code}. The main difference between the C code and its Matlab counterpart is the
level of detail reached by the instructions; while Matlab uses vectors, matrices, and direct operations on these, the C program only performs scalar operations. In addition, some 1-line functions now extend over several lines. It is therefore interesting to see how the proofs developed for the earlier Matlab code apply to this more detailed C code. 

\subsection{Forward analysis}
We will perform only the forward analysis for the sake of brevity. Skipping variable declarations, the first relevant line of code is {\tt 4: x[0]=0}. We may therefore write the triple 

\vspace{3mm}
\begin{tabular}{l}
$\left\{true\right\}$\\
{\tt 4:   x[0] = 0}\\
$\{ x[0] = 0 \}$
\end{tabular}
\vspace{3mm}

Line ${\tt 5: x[1] = 0}$ completes the initialization of $x= \left[\mbox{x[0] x[1]}\right]^T$, leading to the triple

\vspace{3mm}
\begin{tabular}{l}
$\{ x[0] = 0 \}$\\
{\tt 5: x[1] = 0}\\
$\{ x\in {\cal E}_{P} \}$.\\
\end{tabular}
\vspace{3mm}

Doing as precedently, line {\tt 6:  while(1)\{}, together with line {\tt 20:  \}} leads to the set of commented lines

\vspace{3mm}
\begin{tabular}{l}
$\{ x\in {\cal E}_{P} \}$\\
6:   while(1)\{\\
$\{ x\in {\cal E}_{P} \}$\\
$\vdots$ \\
$\{ x\in {\cal E}_{P} \}$\\
{\tt 20:  \}}\\
$\{ false \}$

\end{tabular}
\vspace{3mm}

Line {\tt 7:    fscanf(stdin,"$\backslash$\%f",\&y);} likewise leads to the triple

\vspace{3mm}
\begin{tabular}{l}
$\{ x\in {\cal E}_{P} \}$\\
{\tt 7:    fscanf(stdin,"$\backslash$\%f",\&y);}\\
$\{ x\in {\cal E}_{P} \}$.
\end{tabular}
\vspace{3mm}

The test {\tt 8:   if (y >1)\{} $ \ldots$ {\tt 10: \}} may be commented using insight from line 9 as

\vspace{3mm}
\begin{tabular}{l}
$\{ x\in {\cal E}_{P} \}$\\
{\tt 8:   if (y >1)\{}\\
$\{ x\in {\cal E}_{P}, \; y>1 \}$\\
\vdots\\
$\{ x\in {\cal E}_{P}, \; y=1 \}$\\
{\tt 10:	   \}}\\
$\{ x\in {\cal E}_{P}, \; y<=1 \}$.
\end{tabular}
\vspace{3mm}

which is of course compatible with the assignment {\tt 9:  y = 1} 

\vspace{3mm}
\begin{tabular}{l}
$\{ x\in {\cal E}_{P}, \; y>1 \}$\\
{\tt 9: y = 1}\\
$\{ x\in {\cal E}_{P}, \; y=1 \}$.
\end{tabular}
\vspace{3mm}

The second test {\tt 11:	if (y<-1)\{} $\ldots$ {\tt 13:	   \}} is documented the same way

\vspace{3mm}
\begin{tabular}{l}
$\{ x\in {\cal E}_{P}, \; y<=1 \}$\\
{\tt 11:	if (y<-1)\{}\\
$\{ x\in {\cal E}_{P}, \; y<-1 \}$\\
\vdots\\
$\{ x\in {\cal E}_{P}, \; y=-1 \}$\\
 {\tt 13:	   \}}\\
$\{ x\in {\cal E}_{P}, \; y^2 \leq 1$
\end{tabular}
\vspace{3mm}

and line {\tt 12:      y=-1} yields the triple

\vspace{3mm}
\begin{tabular}{l}
$\{ x\in {\cal E}_{P}, \; y<-1 \}$\\
{\tt 12:      y=-1} \\
$\{ x\in {\cal E}_{P}, \; y=-1 \}$.
\end{tabular}
\vspace{3mm}

The latter postcondition is clearly the same as that of line 7 of 
the Matlab program shown in Fig.~\ref{matlab-code}. Following the 
same process as for the Matlab program, the precondition for line 
{\tt 14:    u = 564.48*x[0]-1280*y} is chosen to be weaker than 
the post-condition of line 13; we choose the precondition
\[
\left\{\left[\begin{array}{l}x\\y\end{array}\right] \in {\cal E}_Q \right\}, \;\; \mbox{ with } 
Q = \left[\begin{array}{cc} 0.01 P & 0 \\ 0 & 0.99 \end{array} \right],
\]
to generate the triple

\vspace{3mm}
\begin{tabular}{l}
$\left\{\left[\begin{array}{l}x\\y\end{array}\right] \in {\cal E}_Q \right\}$\\
{\tt 14:    u = 564.48*x[0]-1280*y}\\
$\left\{\left[\begin{array}{l}x\\y\end{array}\right] \in {\cal E}_Q, \; u^2 \leq \left[C \; D \right]Q^{-1}\left[C\; D\right]^T\right\} $
\end{tabular}
\vspace{3mm}

Line {\tt 15:    fprintf(stdout,"\%f$\backslash$n",u);} may in turn be instrumented as the triple

\vspace{3mm}
\begin{tabular}{l}
$\left\{\left[\begin{array}{l}x\\y\end{array}\right] \in {\cal E}_Q, \; u^2 \leq \left[C \; D \right]Q^{-1}\left[C\; D\right]^T\right\} $\\
15:    fprintf(stdout,"\%f$\backslash$n",u);\\
$\left\{\left[\begin{array}{l}x\\y\end{array}\right] \in {\cal E}_Q \right\}$
\end{tabular}
\vspace{3mm}

Lines 16 and 17 utilise a buffer variable aimed at performing the controller state update. The effect of these lines
is easier to figure out by means of matrix operations: For example, line {\tt 16:    xb[0]=x[0]} can be seen as the matrix operation

\[
\left[\begin{array}{cc} 
x\\
y \\
\mbox{xb}[1]
\end{array} \right] = 
\left[\begin{array}{ccc}
1 & 0 & 0 \\
0 & 1 & 0\\
0 & 0 & 1\\
1 & 0 & 0
\end{array} \right] \left[\begin{array}{c}x \\ y \end{array}\right]
\]

Given this matrix operation, it is easy to figure out the image of the ellipsoid ${\cal E}_Q$ through this mapping. The 
parameterization of the resulting ellipsoid needs changing, however, because it is singular (it is flat along one of its semi-axes). Whenever the matrix $R$ is invertible, the following define the same ellipsoid:
\[
\left\{x \;|\; x^TR^{-1}x \leq 1\right\} = \left\{x \;|\; \left[\begin{array}{cc} 1 & x^T \\ x & R \end{array}\right]\geq 0 \right\}
\]
When $R$ is not invertible, then only one of the statements above defines an ellipsoid. For the sake of conciseness, we will name ${\cal G}_R$ the ellipsoid defined as
\[
{\cal G}_R = \left\{x \;|\; \left[\begin{array}{cc} 1 & x^T \\ x & R \end{array}\right]\geq 0 \right\}.
\]

With such an alternative ellipsoid definition we may write a concise triple for line 16 as

\vspace{3mm}
\begin{tabular}{l}
$\left\{\left[\begin{array}{l}x\\y\end{array}\right] \in {\cal E}_Q \right\}$\\
{\tt 16:    xb[0]=x[0]}\\
$\left\{\left[\begin{array}{l}x\\y\\\mbox{xb}[1]\end{array}\right]\in{\cal G}_R  \right\}$
\end{tabular}
\vspace{3mm}

with 
\[
R = 
\left[\begin{array}{ccc}
1 & 0 & 0 \\
0 & 1 & 0\\
0 & 0 & 1\\
1 & 0 & 0
\end{array} \right]Q^{-1}\left[\begin{array}{ccc}
1 & 0 & 0 \\
0 & 1 & 0\\
0 & 0 & 1\\
1 & 0 & 0
\end{array} \right]^T
\]

Line {\tt 17:    xb[1]=x[1];} may be treated likewise, and yield the triple

\vspace{3mm}
\begin{tabular}{l}
$\left\{\left[\begin{array}{l}x\\y\\\mbox{xb}[1]\end{array}\right] \in {\cal G}_R \right\}$\\
{\tt 17:    xb[1]=x[1];}\\
$\left\{\left[\begin{array}{l}x\\y\\xb\end{array}\right]\in{\cal G}_S  \right\}$
\end{tabular}
\vspace{3mm}

where 
\[
S = 
\left[\begin{array}{cccc}
1 & 0 & 0 & 0\\
0 & 1 & 0 & 0\\
0 & 0 & 1 & 0\\
0 & 0 & 0 & 1 \\
0 & 1 & 0 & 0
\end{array} \right]R\left[\begin{array}{cccc}
1 & 0 & 0 & 0 \\
0 & 1 & 0 & 0\\
0 & 0 & 1 & 0\\
0 & 0 & 0 & 1\\
1 & 0 & 0 & 0
\end{array} \right]^T \mbox{ and } xb = \left[\begin{array}{l}\mbox{xb}[1] \\ \mbox{xb}[2]\end{array} \right].
\]

Continuing along the same lines the triple for {\tt 18:    x[0]:= 0.4990*xb[0]-0.0500*xb[1]+y;}
is 

\vspace{3mm}
\begin{tabular}{l}
$\left\{\left[\begin{array}{l}x\\y\\xb\end{array}\right]\in{\cal G}_S  \right\}$\\
{\tt 18:    x[0]:= 0.4990*xb[0]-0.0500*xb[1]+y;}\\
$\left\{\left[\begin{array}{l}x\\y\\xb\end{array}\right]\in{\cal G}_T  \right\}$\\
\end{tabular}
\vspace{3mm}

with

\[
T = 
\left[\begin{array}{ccccc}
0 & 0 & 1 & 0.4990 & -0.0500\\
0 & 1 & 0 & 0 & 0\\
0 & 0 & 1 & 0 & 0\\
0 & 0 & 0 & 1 & 0\\
0 & 0 & 0 & 0 & 1
\end{array} \right]S\left[\begin{array}{ccccc}
0 & 0 & 1 & 0.4990 & -0.0500\\
0 & 1 & 0 & 0 & 0\\
0 & 0 & 1 & 0 & 0\\
0 & 0 & 0 & 1 & 0\\
0 & 0 & 0 & 0 & 1
\end{array} \right]^T,
\]

and the triple for line {\tt 19:    x[1]:= 0.01*xb[0]+xb[1];} is

\vspace{3mm}
\begin{tabular}{l}
$\left\{\left[\begin{array}{l}x\\y\\xb\end{array}\right]\in{\cal G}_T  \right\}$\\
{\tt 19:    x[1]:= 0.01*xb[0]+xb[1];}\\
$\left\{x\in{\cal G}_U  \right\}$
\end{tabular}
\vspace{3mm}

with 
\[
U = \left[
\begin{array}{ccccc}
1 & 0 & 0 & 0 & 0\\
0 & 0 & 0 & 0.01 & 1
\end{array} \right]T\left[
\begin{array}{ccccc}
1 & 0 & 0 & 0 & 0\\
0 & 0 & 0 & 0.01 & 1
\end{array} \right]^T.
\]

Explicit calculations then show $U^{-1}\geq P$, which ensures consistency between the postcondition of line 19 and the precondition of line 20. Moreover, it is easy to show that 
$U = \tilde{P}^{-1}$, where $\tilde{P}$ is defined in Eq.~(\ref{ptilde}).

\subsection{Remarks}

Bringing the forward analyses of the Matlab code, on the one hand, and of the C code, on the other hand, shows how the higher level-of-detail present in the C code instructions yields considerably more detailed information about the behavior of each variable. This hgher level of detail is precisely what makes the analysis of the C code necessary beyond that of the Matlab code. The C code analysis is, however, not complete and still contains ambiguities. Consider for example Line {\tt 18:    x[0]:= 0.4990*xb[0]-0.0500*xb[1]+y;}. Depending on the compiler used, this line may be executed as {\tt  x[0]:= (0.4990*xb[0]+(-0.0500*xb[1]+y));} or {\tt  x[0]:=((0.4990*xb[0]-0.0500*xb[1])+y);}, for example. The resulting intermediate values may differ widely, yet must be computed in advance to offer a complete analysis of the real time computations that take place during comntroller operations. This justifies a further, more detailed analysis of the code that is produced by compiling that given in Fig.~\ref{code}. 

In addition, it is clear that the analyses shown on a simple second-order controller example easily extend to linear controllers of arbitrary order (possibly subject to input saturation). 

\section{Conclusions}

In the first part of this report, we have advocated the necessity of documenting not only the specifications, but also the implementation of control systems. Using well-known proofs about the behaviors of controller specifications, we have shown how such proofs may be translated as proofs about various levels of implementation of such controllers.


\begin{thebibliography}{CCF{\etalchar{+}}05}

\bibitem[AG64]{AiG:64}
M.~A. Aizerman and F.~R. Gantmacher.
\newblock {\em Absolute stability of regulator systems}.
\newblock Information Systems. Holden-Day, San Francisco, 1964.

\bibitem[AR04]{RV2004ArkoudasAndRinard}
K.~Arkoudas and M.~Rinard.
\newblock Deductive runtime certification.
\newblock In {\em Proceedings of the 2004 Workshop on Runtime Verification (RV
  2004)}, Barcelona, Spain, April 2004.

\bibitem[BEFB94]{BEFB:94}
S.~Boyd, L.~{El~{G}haoui}, E.~Feron, and V.~Balakrishnan.
\newblock {\em Linear Matrix Inequalities in System and Control Theory},
  volume~15 of {\em SIAM Studies in Applied Mathematics}.
\newblock SIAM, 1994.

\bibitem[CCF{\etalchar{+}}05]{CousotEtAl-ESOP05}
P.~Cousot, R.~Cousot, J.~Feret, A.~Min\'{e}, D.~Monniaux, L.~Mauborgne, and
  X.~Rival.
\newblock The {ASTR\'EE} analyzer.
\newblock In {\em In S.~Sagiv (Ed.), Programming Languages and Systems,
  14${}^{\mathit{th}}$ European Symposium on Programming, ESOP 2005, Held as
  Part of the Joint European Conferences on Theory and Practice of Software,
  ETAPS 2005, Lecture Notes in Computer Science 3444}, pages 21--30, Edinburgh,
  UK, April 4--8, 2005, 2005. Springer.

\bibitem[Fer04]{Fer:04}
J.~Feret.
\newblock Static analysis of digital filters.
\newblock In {\em Lecture Notes in Computer Science, Programming Languages and
  Systems}, volume 2986/2004, pages 33--48. Springer-Verlag, 2004.

\bibitem[Flo67]{Flo:67}
R.W. Floyd.
\newblock Assigning meanings to programs.
\newblock In J.~T. Schwartz, editor, {\em Mathematical Aspects of Computer
  Science, Proceedings of Symposia in Applied Mathematics}, volume~19, pages
  19--32, Providence, RI, december 1967. American Mathematical Society.

\bibitem[Hoa69]{Hoa:69}
C.A.R. Hoare.
\newblock An axiomatic basis for computer programming.
\newblock {\em Communications of the ACM}, 12(10):576--583, October 1969.

\bibitem[Mon03]{Mon:03}
J.~F. Monin.
\newblock {\em {S}oftware {R}eliability {M}ethods}.
\newblock Springer, 2003.

\bibitem[Pel01]{Pel:01}
D.~A. Peled.
\newblock {\em {S}oftware {R}eliability {M}ethods}.
\newblock Springer, 2001.

\bibitem[Rin99]{Rin:99}
M.~Rinard.
\newblock Credible compilation.
\newblock Technical Report MIT-LCS-TR-776, Massachusetts Institute of
  Technology, Laboratory for Computer Science, Cambridge, MA, 1999.

\bibitem[RM99]{rinard99credible}
Martin Rinard and Darko Marinov.
\newblock Credible compilation with pointers.
\newblock In {\em FLoC Workshop on Run-Time Result Verification, Trento,
  Italy}, 1999.

\end{thebibliography}

\newcommand{\etalchar}[1]{$^{#1}$}

\end{document}